\documentclass[sigconf]{acmart}
\AtBeginDocument{%
  \providecommand\BibTeX{{%
    \normalfont B\kern-0.5em{\scshape i\kern-0.25em b}\kern-0.8em\TeX}}}
 \usepackage{subcaption}
\usepackage{booktabs}
\usepackage{colortbl}
\usepackage{float}
\usepackage{fancybox}
\usepackage[compact]{titlesec}         
\titlespacing{\section}{0pt}{0pt}{0pt} 
\AtBeginDocument{
  \setlength\abovedisplayskip{0pt}
  \setlength\belowdisplayskip{0pt}}

\usepackage[inline]{enumitem}
\definecolor{gray}{rgb}{0.01,0.199,0.1}

\newcommand{\nd}{\vspace{1mm}\noindent}

\renewcommand{\paragraph}[1]{\noindent\textsf{#1}.}

\copyrightyear{2022}
\acmYear{2022}
\setcopyright{acmcopyright}\acmConference[EASE 2022]{The International Conference on Evaluation and Assessment in Software Engineering 2022}{June 13--15, 2022}{Gothenburg, Sweden}
\acmBooktitle{The International Conference on Evaluation and Assessment in Software Engineering 2022 (EASE 2022), June 13--15, 2022, Gothenburg, Sweden}
\acmPrice{15.00}
\acmDOI{10.1145/3530019.3530039}
\acmISBN{978-1-4503-9613-4/22/06}

\begin{document}

\title{Studying the Practices of Deploying Machine Learning Projects on Docker}

\author{Moses Openja}
\affiliation{%
  \institution{Polytechnique Montreal}
  \city{Montreal}
  \state{Quebec}
  \country{Canada}
  \postcode{43017-6221}
}
\email{openja.moses@polymtl.ca}

\author{Forough Majidi}
\affiliation{%
  \institution{Polytechnique Montreal}
  \city{Montreal}
  \state{Quebec}
  \country{Canada}}
\email{forough.majidi@apolymtl.ca}

\author{Foutse Khomh}
\affiliation{%
  \institution{Polytechnique Montreal}
  \city{Montreal}
  \state{Quebec}
  \country{Canada}
}
\email{foutse.khomh@polymtl.ca}

\author{Bhagya Chembakottu}
\affiliation{%
  \institution{Polytechnique Montreal}
  \city{Montreal}
  \state{Quebec}
  \country{Canada}
}
\email{bhagya.c@polymtl.ca}

\author{Heng Li}
\affiliation{%
  \institution{Polytechnique Montreal}
  \city{Montreal}
  \state{Quebec}
  \country{Canada}
}
\email{heng.li@polymtl.ca}







\begin{abstract}
  Docker is a containerization service that allows for convenient deployment of websites, databases, applications' APIs, and machine learning (ML) models with a few lines of code. Studies have recently explored the use of Docker for deploying general software projects with no specific focus on how Docker is used to deploy ML-based projects. 
  In this study, we conducted an exploratory study to understand how Docker is being used to deploy ML-based projects. As the initial step, we examined the categories of ML-based projects that use Docker. We then examined why and how these projects use Docker, and the characteristics of the resulting Docker images. Our results indicate that six categories of ML-based projects use Docker for deployment, including ML Applications, MLOps/ AIOps, Toolkits, DL Frameworks, Models, and Documentation. We derived the taxonomy of 21 major categories representing the purposes of using Docker, including those specific to models such as model management tasks (e.g., testing, training). We then showed that ML engineers use Docker images mostly to help with the platform portability, such as transferring the software across the operating systems, runtimes such as GPU, and language constraints. However, we also found that more resources may be required to run the Docker images for building ML-based software projects due to the large number of files contained in the image layers with deeply nested directories. We hope to shed light on the emerging practices of deploying ML software projects using containers and highlight aspects that should be improved.
\end{abstract}

\ccsdesc[500]{Machine Learning~Deployment}
\ccsdesc[300]{>Machine Learning~Docker}

\keywords{Machine Learning, Deep Neural Network, Deployment, Docker}

\maketitle

\section{Introduction}\label{sec:introduction}
Docker and its related container technologies have become a prominent solution for automating the deployment of modern software systems. This is true due to its numerous desirable features such as isolation, low overhead, and efficient packaging of the runtime environment. Docker container management framework consists of images containing applications and their required runtime dependencies~\cite{soltesz2007:container} and can be easily versioned, stored, and shared via centralized registry services (e.g., Docker Hub\footnote{https://hub.Docker.com}).

Researchers~\cite{zhao2019large,skourtis2019carving,shu2017study,cito2017empirical,harter2016slacker,brogi2017Dockerfinder} have extensively studied the use of Docker for the deployment process of general software systems. In contrast, we could not find any study focusing on understanding how Docker is being used to deploy machine learning based (ML-based) projects (i.e., Projects using machine learning). 
This information could help the software engineering community understand the emerging practice of deploying ML applications using containers  
and identify aspects that should be improved. 

In this study, we conducted an exploratory empirical study to understand how Docker is being used in the deployment process of ML-based software projects. We analyzed 406 
open-source ML-based software projects that host their source code on GitHub and have their corresponding Docker images hosted on Docker Hub. Specifically, we analyzed the purposes of using Docker, the image information, how Docker is used, 
and the characteristics of Docker images 
based on the following research questions: 

\begin{itemize}
    \item[\textbf{RQ1}] \textbf{What kind of ML-based software projects use Docker?}
    
    This question aims to understand the types of ML-based software projects that use Docker in their deployment process. 
    This information will help us understand if Docker is only being adopted by some specific ML-based software projects or 
    by ML-based software projects in general. Through manual analysis, we grouped the studied ML-based software projects based on their domains and types, into six (6) different categories of `Application System' (42\%), `AIOps' (23\%), `ToolKit' (16\%), `DL Frameworks' (15\%), `Models' (13\%), and ML-based Tutorials/ Documentation (1\%).
    \item[\textbf{RQ2}] \textbf{What is the main purpose of using Docker in ML-based software projects?} The objective of this question is to understand the kinds of automation provided by Docker that are used in the deployment of ML-based software projects. 
    Following an 
    open coding procedure, we analyzed the information related to the created Dockerfiles and Docker images 
    and generated a taxonomy of 21 major categories representing the purposes for using Docker in the deployment process for ML-based software projects, such as Data management, Interactive development, Task scheduling, and Model management. 
    \item[\textbf{RQ3}] \textbf{Which Docker functionalities are used to build ML-based software projects?} This question examines the functionalities of Docker used to build ML-based software projects. We extracted and categorized the information specified within the Dockerfiles such as the types of the base images and the used execution instructions. 
    Our findings show that the RUN command are the most used Docker command to manage File system, dependencies, Permissions, Build/ Execution, 
    and the environment settings, out of which File system and Dependencies related commands are the most executed instructions when building Docker images for ML-based software projects. 
    Similarly, we find that most base images used to build ML software projects are related to operating system (Ubuntu, Debian), platform runtime (e.g., Cuda), language runtime (e.g., Python, Node.js), and machine learning platform (e.g., DL frameworks, AutoML or online ML platform). 
     
    
    
     \item[\textbf{RQ4}] \textbf{What are the characteristics of Docker images used for deploying ML-based software projects?}
    We extracted and analyzed the real Docker images of the studied ML-based software projects from the Docker Hub 
    registry and characterized them based on the information contained in the image manifest files and the image layers such as the target hardware environment, the layers, and the file composition and memory consumption of Docker images used to build ML-based software projects. We observed that the images are built to target different hardware architecture and operating systems but are subsequently associated with larger files that likely  expensive in terms of computation resources. Moreover, we observe that a small set of the contained files in the image layers occupy relatively large image space, which call for more efficient approaches (e.g., using efficient compression) to store these large files. 
\end{itemize}

\noindent\textbf{Paper organization.}
The rest of this paper is organized as follows: 
In Section~\ref{sec:background}, we introduce  Docker-related concepts relevant to our study. Section~\ref{sec:methodology} describes the methodology of our study. In Section~\ref{sec:results}, we report the results of our analysis answering the four proposed research questions. 
Section~\ref{sec:related-works} discusses the related literature while Section~\ref{sec:threats} introduces potential threats to the validity of our results. Finally, in Section~\ref{sec:conclusion}, we further discuss the results, conclude our study and outlines avenues for future works.

\section{Background}\label{sec:background}
\subsection{Docker}\label{subsec:Docker}

Docker and Docker containers allow for packaging an application with its dependencies and execution environment into a standardized, deployable unit, ensuring that the application performs reliably and consistently on the different computing platforms. Docker ecosystem consists of multiple components, including a docker client to allow the user to interact with a running docker daemon. The docker daemon runs a container from a local image or pulls an image directly from the registry.


\subsection{Dockerfile}\label{subsec:Dockerfiles}

According to the definition from Docker documentation~\cite{Dockerfiles-reference:2021}, ``A Dockerfile is a text document that contains all the commands (instructions) a user could call on the command line to assemble an image. Using Docker build, users can create an automated build that executes several command-line instructions in succession''. Dockerfile normally contains details about the base image, environmental variables, comments, and commands to execute shell commands, install dependencies, install software such as compiling and linking, open/ expose ports for external access, and start the process. 

\nd $\bullet$ \textbf{Base Image:} Is the initial point to understand what the project is using Docker for. They specify the base environment (e.g., an operating system) where the Docker image is build on. 
Usually, a base image specification is represented as a tuple of the format (namespace/)image\_name(: version). 
A `image\_name' is used to identify an image and often indicates the image's content. For the `official' images, for example, ubuntu or python, 
the image\_name is the sole identifier of an image. Non-official images further depend on a namespace, which is often the organization's name or user who maintains the image (e.g., \texttt{nvidia/cuda}, \texttt{tensorflow/tensorflow}). Moreover, a base image specification can contain a string version, representing the specific version number (such as 1.0.0), specific target runtime hardware (e.g.-runtime-ubuntu20.04) or a more flexible version (like latest).
    
\nd $\bullet$ \textbf{Docker Instructions:} Are specified within a Dockerfile and are used by Docker for automatically building an image. Usually, they indicate how a given project is built on a base 
image. A Dockerfile can contain all the commands a user could call on the command line to assemble an image. For example, a RUN instruction to execute any commands or a COPY instruction to copy files or directories to the container's filesystem at the path destination location.

\subsection{Images Manifest and Layers}\label{subsec:Dockerimages}

\nd  Docker uses a concept of Docker image (also called container image) to build and package, distribute and run the software application. Docker images may consist of multiple layers representing specific components/dependencies of the Docker image, such as a package or a library. Moreover, the layers may be shared across images that depend on the same elements or components. Any change on the Docker images is reflected inside image layers, 
and the image summary is captured in the image manifest~\cite{Docker-Manifest:V1,Docker-Manifest:V2} file. An image manifest contains a list of layer identifiers (also called digest) for all the layers required by a given image and other descriptions of various parameters of the Docker images (e.g., the environment variables) and the target platform information (e.g., the operating systems and the architecture).

\subsection{Docker Hub}
A Docker Hub is a Docker registry for storing docker images. It is known as one of the most popular registry for storing private and public images. The images are stored in Docker Hub in repositories, each containing different versions of a similar image. The image manifests are stored as a JSON file, while the layers are stored in a compressed format. Users can upload, search or download images. The images provided by Docker Inc or partners are called official images and have a sole name of the form <image\_name>, while the user uploaded images are contained in the repository in the format <username>/<image\_name>.

\section{Study Design}\label{sec:methodology}
\begin{figure}[ht!]
\center
\includegraphics[width=\linewidth]{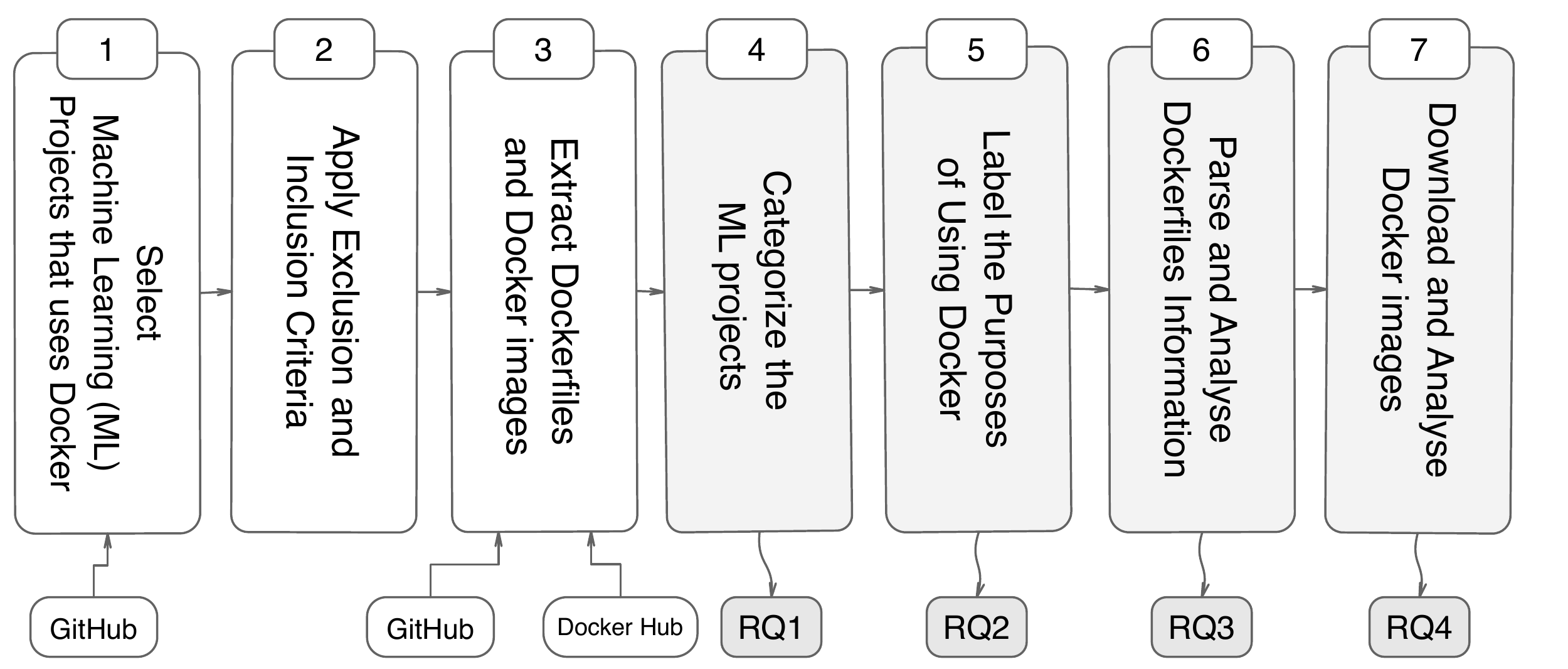}
\caption{An overview of our study methodology.}
\label{fig:methodology}
\end{figure}

This section describes the methodology that we used to conduct this study. We followed a mixture of qualitative and quantitative (sequential mixed-methods~\cite{ivankova:2006:using}) to answer the proposed research questions.  An overview of our methodology is shown in Figure~\ref{fig:methodology}.

\nd In the following, we elaborate on each of the step: 

\nd \textbf{{\Large \textcircled{\normalsize 1}} Select Machine Learning (ML) Projects} 

\nd To select ML-based software projects for our study, we first generated a set of relevant keywords for searching on GitHub. 
Using the GitHub search API~\cite{GitHubSearchAPI:2021}, we search on GitHub for the repositories using the keywords. 
Specifically, we proceed as follows: 

\begin{enumerate}
    \item \emph{Generating the Search Keywords}: This step aims to identify a rich set of keywords (topics) allowing us to capture a broad range of domains of the ML software systems hosted on GitHub. We first searched through GitHub topics with keywords ``machine-learning'', ``deep-learning'', and ``reinforcement-learning'' using the GitHub search API~\cite{GitHubSearchAPI:2021}. Topics are sets of labels assigned to repositories hosted in GitHub to allow searching or exploring through the GitHub repositories basing on the project's types, category or technology. Using these initial keywords returned a set of repositories and their respective topics 
    which we then extracted the resulting topics and manually summarized the topics into $14$ 
    major tag categories ($T$) including \emph{`machine-learning', `deep-learning', `deep-neural-network', `reinforcement-learning, `artificial -intelligence', `computer-vision', `image-processing, `neural-network, `image-classification', `convolutional-neural-networks', `object-detection', `machine-intelligence', `autonomous-vehicles', and `autonomous-driving'}.
    
    \item \textit{Extract Machine Learning Repositories Using $T$}: We queried the GitHub API~\cite{github:api} using the list of keywords obtained in the previous step, searching for repositories that: 1) contain at least one of the keywords in $T$ (case insensitive) either in the repository name, repository labels/ topic, descriptions, or README file; 2) not a forked repository (i.e., are mainline software projects). 
    This search returned a total of $30,139$ unique repositories. 
    
    \item \textit{Filtering Machine Learning Repositories Using Docker}: 
    The goal of our study is to understand the practices of using Docker in the ML-based software project through analysis of the Docker-related information. To this end, we identified the ML-based software projects that use Docker in their deployment process. We checked the presence of the keyword `Dockerfile'  
    (case insensitive) in the file names 
    associated with the latest version of each repository and filtered out the repositories that do not contain any Dockerfiles (i.e., Dockerfile). After this step we remained with $3,075$ ML-based software projects that contains at-least one Dockerfile.

\end{enumerate}

\nd \textbf{{\Large \textcircled{\normalsize 2}} Apply Inclusion/ Exclusion Criteria:}
 
 \nd Following the idea from the related works ~\cite{munaiah:2017:curating,Businge:2018:ICSME,Businge:2019:SANER,businge2022:reuse}, we selected the ML repositories that contain at least 100 commits, at least one fork (to reduce the chance of selecting a student's class assignment), stared at least once and contain at least one release. These criteria allows us to select ML-based software projects that are mature and used by the end-users. This step removed 2,104 repositories and we remained with 971 repositories.
 
\nd \textbf{{\Large \textcircled{\normalsize 3}} Extraction of Dockerfiles and Docker images:}

\nd This step extracts the Dockerfiles from the GitHub repositories of the selected ML-based software projects and the respective Docker images hosted on Docker Hub.
``Dockerfile'' itself is a universal term which we used to search (case insentive) for that filename from the latest version of each of the selected ML project's GitHub repository. Note that, in some cases `dockerfile' is used as the file extension, e.g., \texttt{base.aarch64.dockerfile}, \texttt{runtime.x86\_64.dockerfile.sample}. Also, a project may have many different Dockerfiles 
for their different versions, and different specifications at different levels of the folder (not necessarily in the root directory). We downloaded all the Dockerfiles 
using an automated script written in Python. 

\nd For Docker images we first search for the images with the similar names as the GitHub repository names. Then we manually checked and extracted the Docker images in Docker Hub corresponding to the selected ML-based software projects that returned false results in the first search. In some cases, the Docker image has different name as of the repositories or there are multiple images present with the same repository name, thus we manually looked at the content of Dockerfiles and compare with the image information on Docker Hub. Moreover, in most cases the links to the GitHub repository are included within the image descriptions. We also encountered repositories with Dockerfiles that do not have any corresponding image in the Docker Hub. For such cases, we consider them as unknown and did not included them in the scope of this paper. 

After this step, we remained with $406$ ML-based software projects that host their respective Docker images on Docker Hub.

\nd \textbf{{\Large \textcircled{\normalsize 4}} Categorization of the ML-based software projects}: 

\nd This step analyses the different ML-based software projects that uses Docker in their deployment process. 
We started by looking into repositories and manually labelling the categories of the ML-based software projects. The primary reference of our analysis was the description provided in the repositories and the tags associated with the repositories. For example a repository is  labelled as ToolKit or AutoML in the case where the description is about the set of tools, libraries, resources, or routines that are incorporated into ML-based software projects to streamline the development of ML-based software projects.

Three graduate students with both industry and research backgrounds on software engineering for Machine learning and Release engineering did the labelling. The labelling was done in parallel, and the labels which were not agreed upon  underwent further discussion 
until a consensus was achieved. 
The labeling process resulted in 
six categories of ML software projects including: Application System, DL Framework, AIOps applications,  
ToolKit,  Tutorials/Documentation, and Model. The results for step 4 answer our research question \textbf{RQ1} and are presented in Section~\ref{subsection:ml-category}.

\nd \textbf{{\Large \textcircled{\normalsize 5}} Label the purpose of Using Docker}: 

\nd In this step, we try to understand why Docker is used by  
ML-based software projects. Hence, since the answer resides in the descriptions associated with the Dockerfiles - the heart of the Dockerized repositories or the descriptions of the Docker image (Docker Hub) 
However, 
most of the repositories have more than one Dockerfile associated with 
the purpose of they are being used. In most cases, the Dockerfile 
or the directory where it resides is labelled with its purpose. For example, a Dockerfile residing with the test files (inside a folder labelled tests) is likely related to Dockerfile for testing. For example a Dockerfile with the path: \texttt{`/e2e\_tests/e2eTest.Dockerfile'}~\footnote{\url{https://github.com/catboost/catboost/tree/master/catboost/node-package/e2e_tests}} (extracted from the ML software project \texttt{`catboost/catboost'} under the ToolKit category) is likely related to end-to-end testing. Similarly a Dockerfile with the path: \texttt{ `/Docker/Dockerfile-inference'}~\footnote{\url{https://github.com/blue-oil/blueoil/tree/master/Docker}} (from project:  \texttt{`blue-oil/blueoil'} categoried as MLOps/ AIOps) 
is likely used for inference purposes. 

Moreover, we also 
look into the Dockerfiles and validated the purposes of the files. 
All the manual labelling we did in this section was followed by a similar three-person evaluation followed by a discussion on the mismatched labels 
and finalization of the same as mentioned in the previous step. The results for step 5 answer our \textbf{RQ2} and are presented in Section~\ref{rq2}.

\nd \textbf{{\Large \textcircled{\normalsize 6}} Extract and Analyse the Dockerfile Information:} 

\nd Dockerfile has a standard and pervasive way of writing among developers. However, understanding the categories of instructions being used is the first step to understand the characteristics of the Dockerfile. 
In this step  we first downloaded all the Dockerfiles using an automated script written in Python (included in the replication). 

\nd Next, we parsed and analysed the Dockerfiles information such as the instructions and the base image. To parse the Dockerfile, we used a Python script forked from the Docker parser~\cite{Docker-parser:2021} that parse the Dockerfile using the regular expression and returns a set of key value pairs containing the instructions and the corresponding execution command.

\nd We classified the instructions and the commands being executed following the idea from the previous work by Cito et al~\cite{cito2017empirical}. For example the instruction running a commands such as \{'set', 'export', 'source', 'virtualenv'\} where classified as Environment, \{'chmod', 'chown', 'useradd', 'groupadd', 'adduser', 'usermod', 'addgroup'\} as Permision. 
The results for step 6 answer our \textbf{RQ3} and are presented in details in Section~\ref{subsec:rq3}.




\nd \textbf{{\Large \textcircled{\normalsize 7}} Extract and analyze Docker images from Docker Hub:}

\nd This step analyzes the latest version of the images from Docker Hub. To extract the images of the repositories that was considered in this study, we use an open-source tool called skopeo~\cite{Skopeo}. Skopeo helps us copy all image manifest, the configuration files, and the image layers (as a compressed folder) containing all the files executed by the image. The manifest file is a JSON file, and we can consider it as the heart of the Docker images. It contains a summary of all the layers associated with the image. Similarly, a configuration file is a JSON 
like text file containing the configuration information of the images such as the target hardware architecture, the operating systems, the environment variables, and the history of the Docker commands executed when the image was being build. To analyse the image layers files, we must first decompress them. 
We analyzed the files sizes, the depth of the files residing in the images, and each file type. We summarised the characteristics of the Docker images to answer our \textbf{RQ4} and the results are presented in Section~\ref{rq4}. 

\section{Results}\label{sec:results}

\subsection{\textbf{RQ1: Categories of ML Software   Projects}}\label{subsection:ml-category}
The first stage of our research in this paper is to understand the categories of the software projects. As we concluded in our introduction, the usage of ML-based approaches in the projects has increased recently with the promising results provided by the advanced algorithms. However, the challenges of dependencies with larger projects are also more. The shift to Dockerised ML-software projects was not subtle. 
In order to study the deployment of ML projects on Docker, we first investigate the categories of ML projects that leverage Docker deployment. 
In this RQ, we looked into the studied projects and their details to understand the categories of projects that widely adopt the Dockerisation and which are 
still in the migration process. This study helped us conclude the depth or width 
of practice of using Dockerisation in the ML-based software facets. 

\begin{table*}[t]
    \centering
    \small
    \caption{Summary of the categories of ML software repositories hosted on Github and are using Docker in the deployment process.
    \emph{Category:} ML-based software projects categories, 
    \emph{Repos:} The total number and percentage of repositories associated with the category. 
    \emph{Size:} Median size (LOC) of the repositories,  \emph{Commits:} Median number of commits. 
    \emph{Contrib:} Median number of contributors/ developers of the projects.
    \emph{Forks:} The median of number of forks associated to the repositories. 
    \emph{Stars:} The median number of Stars given in the repositories.
    \emph{Releases:} The median number of Releases.
    \emph{Dockerfiles:} The median number of Dockerfiles associated with the repositories. }
    \label{tab:image-summary}
   
    \resizebox{\textwidth}{!}{\begin{tabular}{m{2.3cm} | c c c c c  c c c m{12cm}}\toprule

     \rowcolor{gray!15}
    \textbf{Category}&	\textbf{Repos (\%)}&	\textbf{Size}&	\textbf{Commits}&	\textbf{Contrib}&	\textbf{Forks}&	\textbf{Stars}&\textbf{Releases}&\textbf{Dockerfiles}&\textbf{Descriptions}\\\midrule
    Model&	13 (3\%)&	11,962&	353&	32&	237&	1,026&	8&	1&This category represents the repositories that host ML models: the collection of artifacts trained using data to perform a given task (e.g., classification, regression, dimensionality reduction, etc.)\\
     
 \rowcolor{gray!10}
Application System	&170 (42\%)&	43,713&	793&	14&	87&	348&	12&	2&These are software programs that use machine learning or deep learning to perform specific tasks for the end-user. The repositories in this category contains at least one ML models integrated with the rest of the code or components (e.g., user interface, detection system) to perform specific task such as recommendation, autonomous driving.\\

 MLOps/ AIOps&	92 (23\%)&	24,565&	1,145&	22&	107&	422&	19&	4&Machine Learning Operations (MLOps) are software that helps simplify the continuous management, logistics, and deployment of machine learning models by bridging the gaps between the operations teams and machine learning researchers. This category also combines the AIOps software (whose main functions include monitoring performance, event analysis, correlation, and IT automation) and DataOps. 
 \\

 \rowcolor{gray!10}
Toolkit&	63 (16\%)&	20,937&	1,153&	25	&193&	1,395&	14&	2&This category consists of a set of tools, snippets, plugins, libraries, resources, or routines incorporated into ML-based software applications to streamline the development of ML-based software projects. The category also include Automated Machine Learning (AutoML) libraries that automate the building of scalable machine learning models for a given real-world problem. Examples in our list include Auto-PyTorch, nni, auto-sklearn, autokeras, optuna, AutoBazaar.\\

 DL Framework&	62 (15\%)&	47,961&	1,620&	36&	216&	1,028&	15&	3& These ML-based software project expose a set of routines, functions, algorithms, or applications to provide the building blocks for designing, training, and validating deep neural networks using a high-level programming interface. Examples include PyTorch, MXNet, TensorFlow, and those that rely on GPU-accelerated libraries, such as DALI, cuDNN, and NCCL to deliver high performance, multi-GPU accelerated training.
 \\

 \rowcolor{gray!10}
 Documentations&	6 (1\%)&	87,791&	828&	39&	144&	677&	3&	4&Repositories related to documentation, tutorials or course materials on Machine learning\\
\bottomrule
 \rowcolor{gray!15}
&\textbf{406}&&&&&&&&\\
\bottomrule
    
    \end{tabular}}

\end{table*}


\nd Table~\ref{tab:image-summary} summarises the results of our categorization as per step~{\Large \textcircled{\normalsize 4}} into six ($6$) classes agreed by the two reviewers of the dataset. Most ML software projects are related to the Application System category of ML-based software projects, and the applications tend to use Dockerisation in their practice a lot to improve the portability of their project (e.g., exposing the API for inference or hosting the model within the Docker container storage, with fewer lines of code). AIOps and ML based ToolKit are the next two categories (with 23\% and 16\% respectively) introducing the Docker component in their pipeline. We observed the repositories belonging to Documentation category (e.g., \texttt{`JdeRobot/RoboticsAcademy'}) with the least number of projects. 
Moreover, the median number of Dockerfiles associated with each category is more than one (inclusive of Documentation) except for the Model category. 
It shows that developers tend to use the Dockerisation concept for different purposes, leading to the second research question. The further details mentioned in the table validate that the projects we are considering for our given study are not puppet projects and the inclusion-exclusion criteria we adopted in formulating our dataset are valid.

\begin{center}
\fbox{\parbox{0.45\textwidth}{
We observed that a variety of ML-based software projects use Docker for deployment, including ML Applications System, MLOps/ AIOps projects, Toolkits, DL Frameworks, ML models, and Documentation repositories. Moreover, one project often uses more than one Dockerfile, indicating that these projects use Docker for multiple purposes.
}}
\end{center}

\subsection{\textbf{RQ2: The Purposes of using Docker}}\label{rq2}

\nd This subsection reports our discovered purposes of using Docker in ML-based software projects. 

\begin{figure}[ht!]
\center
\includegraphics[width=1\linewidth]{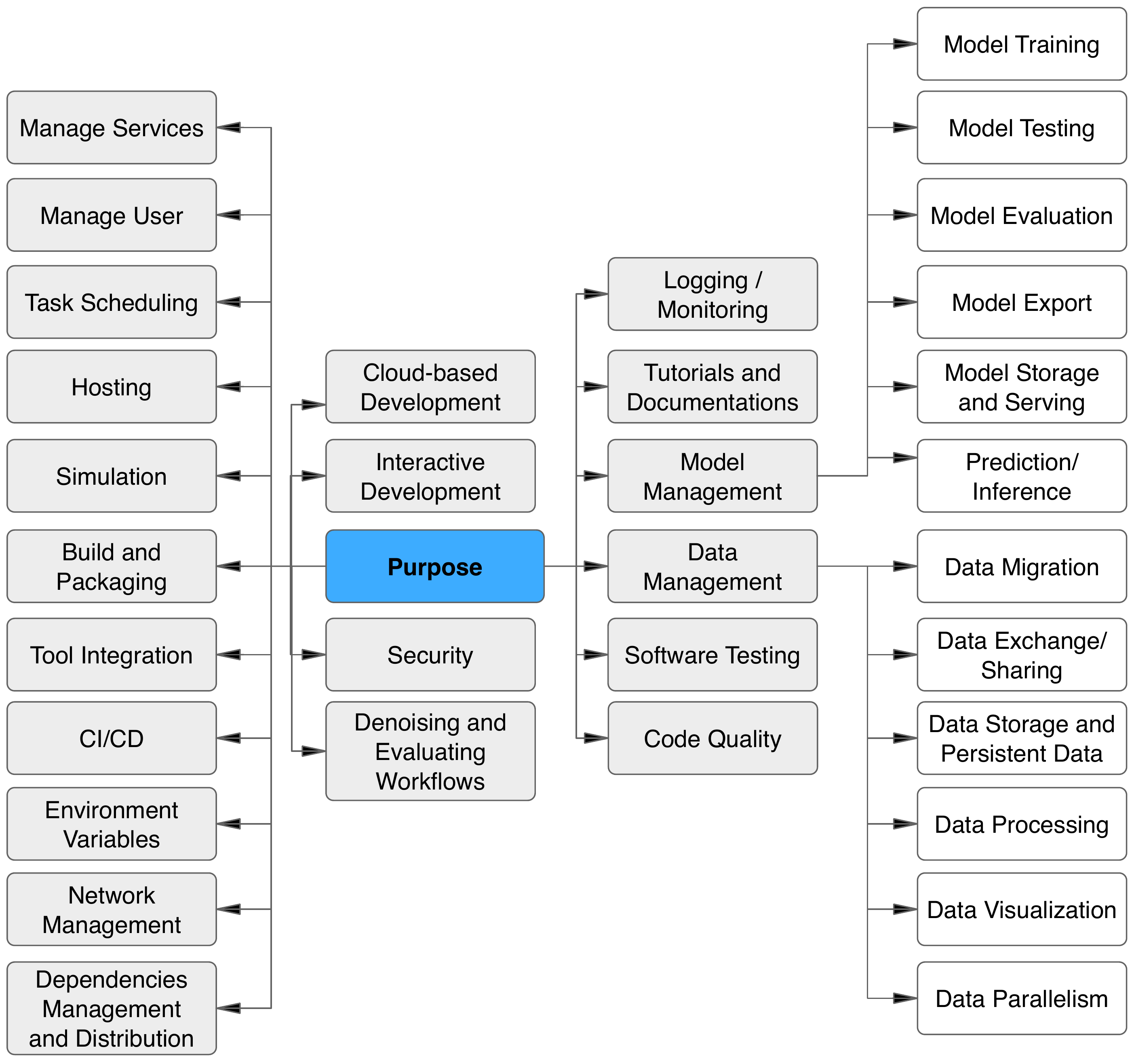}
\caption{Overview of the purposes of using Docker in the studied ML-based  software projects. The 21 
high-level categories of the purposes are highlighted in light gray color code, while the sub-categories are shown in white boxes.}
\label{fig:purpose}
\vspace{-15pt}
\end{figure}

\nd Figure \ref{fig:purpose} presents $21$ 
high-level categories of the purposes of using Docker (in light grey color) observed in the studied ML-based software projects (identified in Step~{\Large \textcircled{\normalsize 5}} of our analysis methodology Section~\ref{sec:methodology}). The purposes reported on the left side of Figure \ref{fig:purpose} are more general, while those on the right are more related to ML components. In the following, we describe some of the categories and sub-categories of the obtained taxonomy in details, highlighting the examples.

\nd $\bullet$ \textbf{\emph{Logging and Monitoring:}} This category is about using Docker to identify suspicious activities in any privileged operating of ML-based  software projects through logging and monitoring. Logging includes tracking and storing records related to the evens, data input, processes, data output, and final results of the running ML-based  software projects. On the other hand, monitoring is a diagnostic tool used for alerting ML engineers (visualization) of the related issues by analyzing metrics. Usually, Logging and monitoring help ensure application availability and assess the impact of state transformations on performance.

\nd $\bullet$ \textbf{\emph{Cloud-based Development:}} This category is about using Docker to automate the process of setting the software development environment on the server-side (i.e., accessible through the browser) connected to cloud-based infrastructure (e.g., CI/CD and version-controlled system~\cite{openja2020:analysis,fylaktopoulos2016:overview,combe2016:docker,openja2021:empirical}) and other services such as a database. More specifically Docker is used in the setting up of the workspace that is ready-to-code 
where all the dependencies needed by the source code are compiled, running build tools and automated testing (e.g., on git push), and live sharing of code.
    
\nd $\bullet$ \textbf{\emph{Interactive Development:}} This category encompasses the use of Docker for deploying and distributing the interactive development environment such as Jupiter notebook or RStudio that allows data scientists or ML Engineers to create and share documents that integrate live code, equations, visualizations, computational output, and other multimedia resources.

\nd $\bullet$ \textbf{\emph{Model management:}} In this category, Docker is used to handle the different activities related to the trained ML algorithms (ML model) that can generate predictions using patterns in the input data. This process includes managing data flow to ML models, working with multiple models, and collecting and analyzing metrics throughout the life cycle of models. For example, in Figure~\ref{fig:purpose} we highlight the different scenarios where Docker is used when working with ML models in the studied ML-based  software projects, such as exposing the API for prediction/ inference, using Docker for Model training/ testing, Model export, model evaluation, and model storage.
       
    
    

    
\nd $\bullet$ \textbf{\emph{Manage Users:}} In this category, the ML engineers use Docker to create and manage users, such as granting varying levels of access based on requirements.
    
    

    
    
    
\nd $\bullet$ \textbf{\emph{CI/CD:}} This category combined using Docker to automate the continuous integration and delivery or deployment of ML-based  software projects, usually to bridge the gaps between development, teams, and operation activities by enforcing automation in building, testing, and deploying ML-based  software projects. 
    
    
        
\begin{center}
\fbox{\parbox{0.45\textwidth}{

There is a broader range of 21 
major purposes of using Docker in the deployment process of ML-based software projects, such as model management, software testing, setting interactive development, data management, 
checking the code quality, distribution of tutorials/ documentation, 
and build/ packaging among others. ML engineers can learn from our observations and implement Docker in their deployment process for these similar purposes. 
}}
\end{center}



\subsection{RQ3: The Used Functionalities Provided by Docker} \label{subsec:rq3}

This section presents the classification results of Docker instructions and base images specified in the Dockerfiles of the selected ML software to understand the Docker functionality used to build the images for ML software projects.
\subsubsection{\textbf{Composition of Docker Instructions}}




\begin{figure}
     \centering
     \begin{subfigure}[b]{\linewidth}
         \centering
         \includegraphics[width=\linewidth]{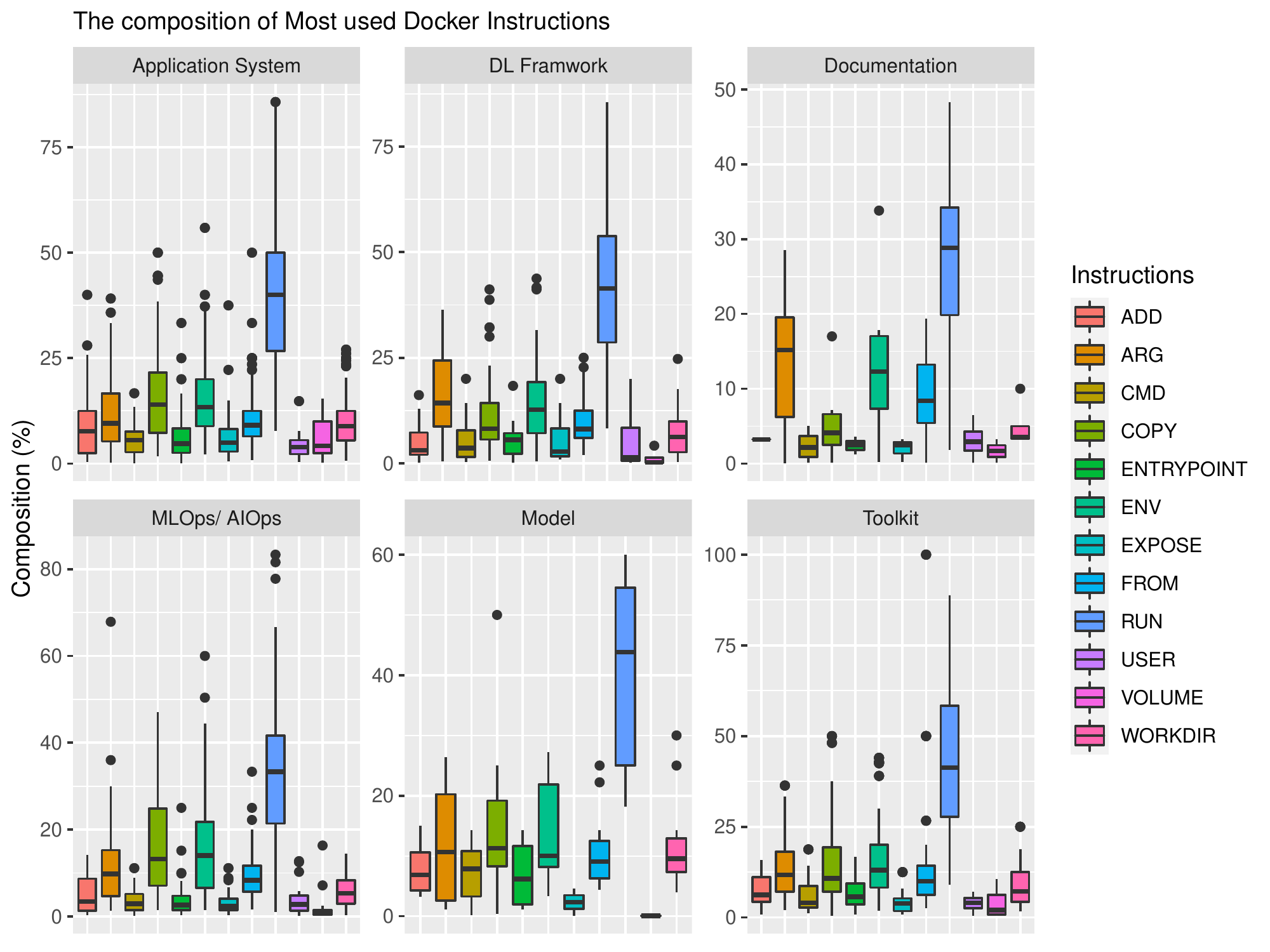}
         \caption{The composition of commonly used Docker instructions across the Dockerfiles of the studied ML-based software projects.}
         \label{fig:instructions-composition}
     \end{subfigure}
     \hfill
     \begin{subfigure}[b]{\linewidth}
         \centering
         \includegraphics[width=\linewidth]{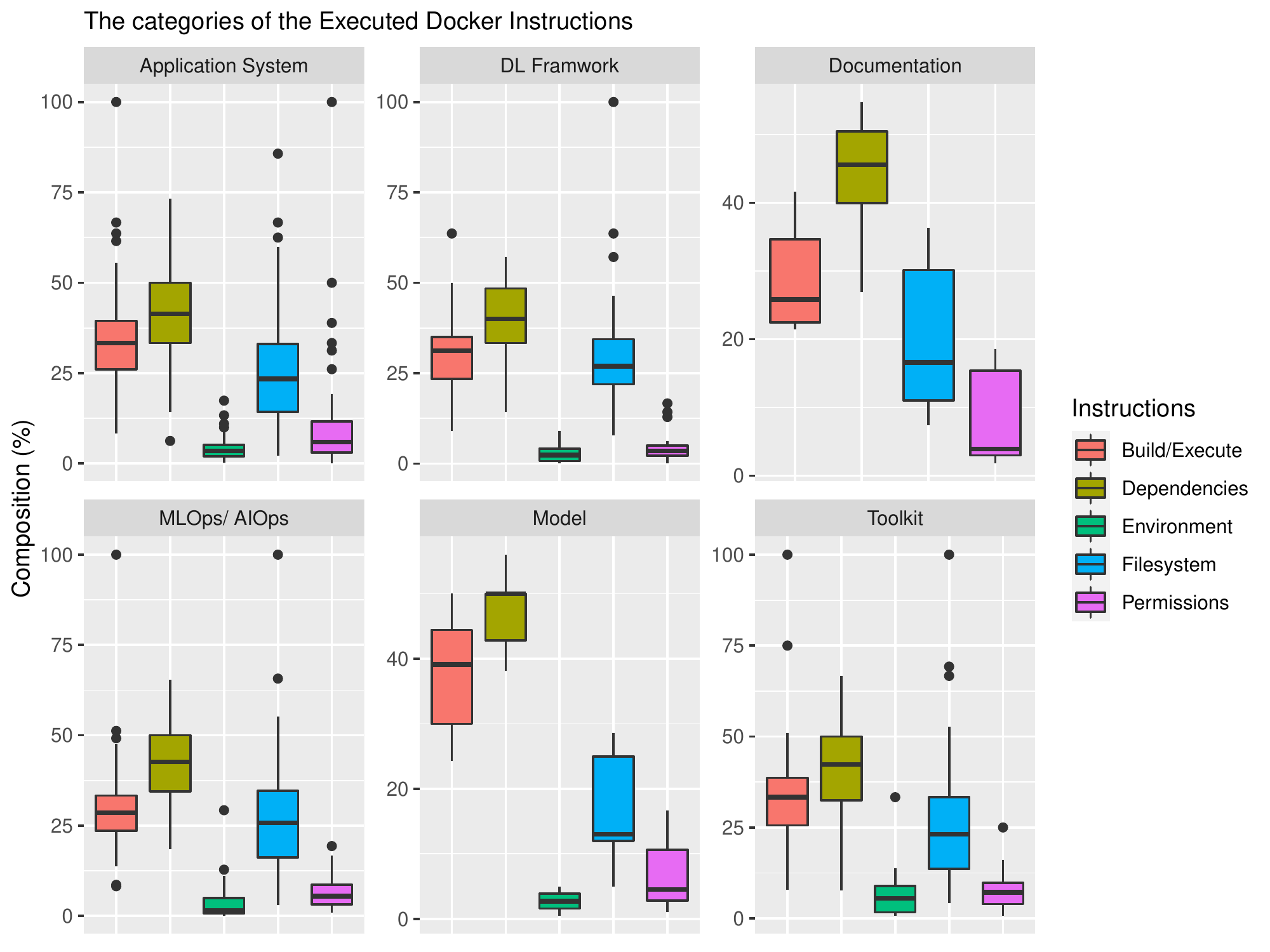}
         \caption{The categories of the commands specified with the RUN instruction 
         representing the main use for the Docker in ML-based  software projects.}
         \label{fig:Docker-category}
     \end{subfigure}
        \caption{The percentage composition of the most used Docker execution command (Instructions) and the categories representing the main use of Docker in ML-based  software projects.}
        \Description{The percentage composition of the most used Docker execution command (Instructions) and the categories representing the main use of Docker in ML-based  software projects.}
        \label{fig:Docker-Instructions}
        \vspace{-15pt}
\end{figure}

\nd Figure~\ref{fig:instructions-composition} presents the percentage breakdown of the commonly used Docker instructions in the Dockerfiles of the studied ML-based  software projects. We did not present the results for Docker instructions (such as LABEL, SHELL, MAINTAINER) that constitute less than 0.1\% percentage composition across at least 50\% of the studied ML-based  software projects. From Figure~\ref{fig:instructions-composition}, we can observe that the RUN is the far most used Docker instruction in Dockerfiles compositing about 40\% of Docker instructions in Dockerfile. This is not surprising given the generic nature of the RUN instruction that allows execution of any viable (non-interactive) shell command within the container. 
Other Docker instructions used in many Dockerfiles but represent smaller fractions of Docker code are: ADD, CMD, COPY, ENTRYPOINT, EXPOSE, USER and VOLUME. The EXPOSE instruction indicates the ports a container should listen to during connections, usually external access, such as endpoints/ API calls. The VOLUME instruction is used to expose any data storage area (database or configuration storage), or folders/ files system created by the Docker container. The reader may refer to~\cite{Dockerfiles-practice:2021} for further details of each instruction. Also, by looking at the usage of instructions within ML-based software project's categories, the outliers of the mandatory FROM command in most of the categories (e.g., in DL Framework, Model, Application System) indicate  
that  
some of the ML software projects prefer to specify multiple FROM instructions within the same Dockerfile compared to using only one FROM instruction. 
The use of multiple FROM instructions within a single Dockerfile allows for creating multiple images or uses one build stage as a dependency for another. In addition, each FROM instruction clears any state created by previous instructions.

Looking at the high percentage composition of the RUN instruction, we provide a further breakdown of what kinds of commands are being executed most in the studied ML-based software projects. Figure~\ref{fig:Docker-category} provides the breakdown of the most used categories of commands being specified using RUN instructions. Specifically, we reused the categories provided by Cito et al~\cite{cito2017empirical} to group the different commands executed by the RUN instructions into five major categories (i.e., Dependencies, Filesystem, User permission, Build/ Execution, and Environment). In the following, we summarised each of the categories:
\begin{enumerate*}
    \item Dependencies: this category is related to commands for managing packages within the Docker or builds commands, such as pip, apt-get, install.
    \item Filesystem: represents difference UNIX utilities used to interact with the file system, such as mkdir or cd.
    \item Permissions: are the UNIX utilities and commands used for managing the user permission when operating with Docker. For example, such as using chmod to change the access permissions of file system objects.
    \item Build/Execution: build tools such as make.
    \item Environment: UNIX commands that set up the correct environment, such as set or source. 
    
 \end{enumerate*}
 

From Figure~\ref{fig:Docker-category} we can see that the dependencies, build/ execution and the filesystems are the three commonly used RUN commands executed. A possible explanation for a high percentage of Filesystem and dependencies related commands is that ML software projects are associated with multiple files and libraries or packages that need to be integrated into the application stack. Introducing an efficient procedure of defining the dependencies and Filesystems may help minimize the complexity of Dockerfile and improve their maintainability.

\begin{center}
\fbox{\parbox{0.45\textwidth}{
The RUN instruction is the most used Docker instruction in all categories of ML projects. In particular, ML projects use the RUN instruction to manage File systems, Dependencies, User Access/ Permission, Build/ Execution, and Environment settings. 
}}
\end{center}


\subsubsection{\textbf{Composition of Base Image}}

As mentioned in Section~\ref{sec:background}, the base image is the initial point to understand what the project is using Docker for. This SubSection reports the base images and the base image types used to build the Docker images for deploying ML-based software projects. 

\begin{figure}
     \centering
     \begin{subfigure}[b]{\linewidth}
         \centering
         \includegraphics[width=\linewidth]{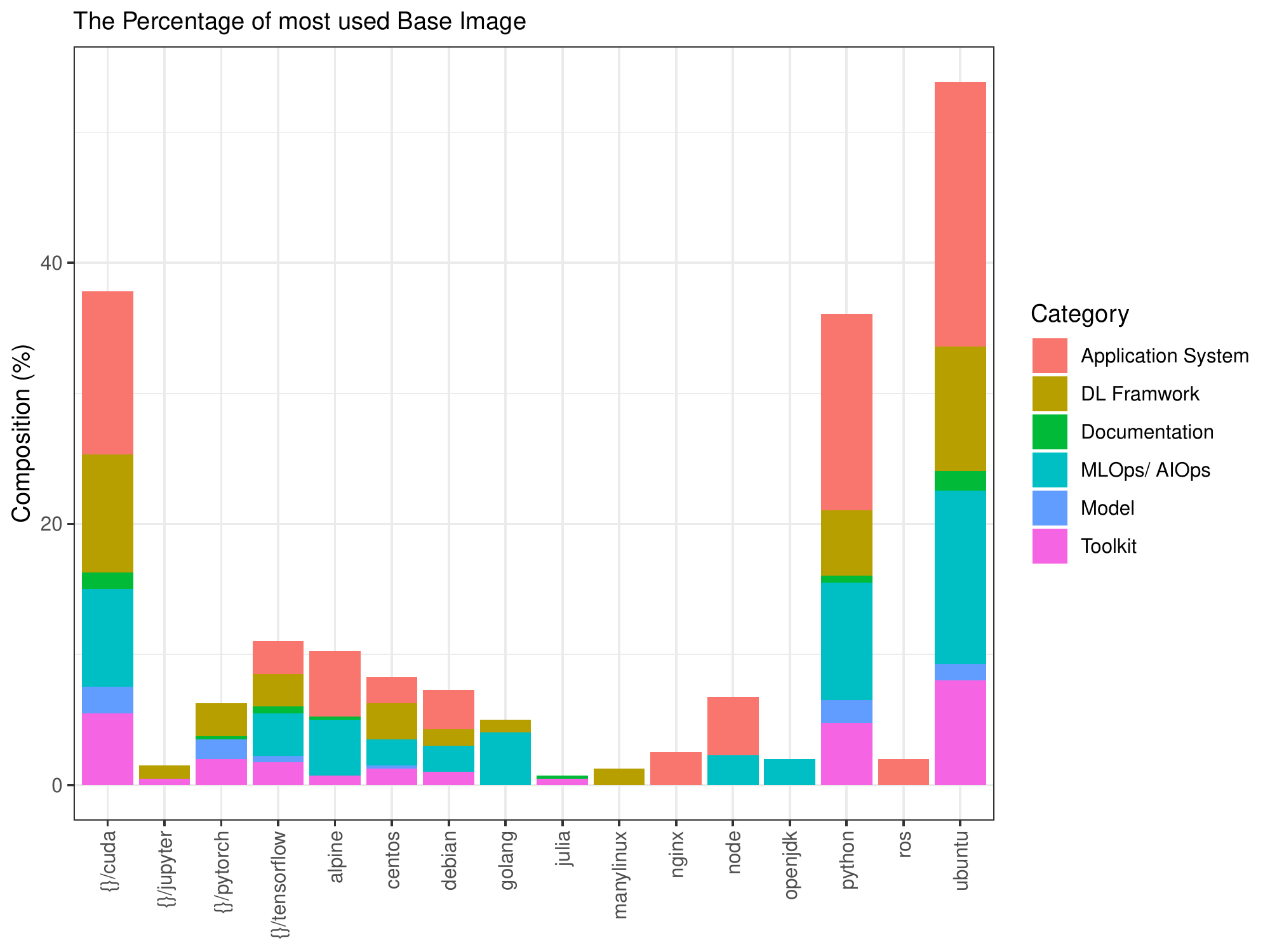}
         \caption{The percentage distribution 
         of the most commonly used Base Images.}
         \label{fig:Base_image}
     \end{subfigure}
     \hfill
     \begin{subfigure}[b]{\linewidth}
         \centering
         \includegraphics[width=\linewidth]{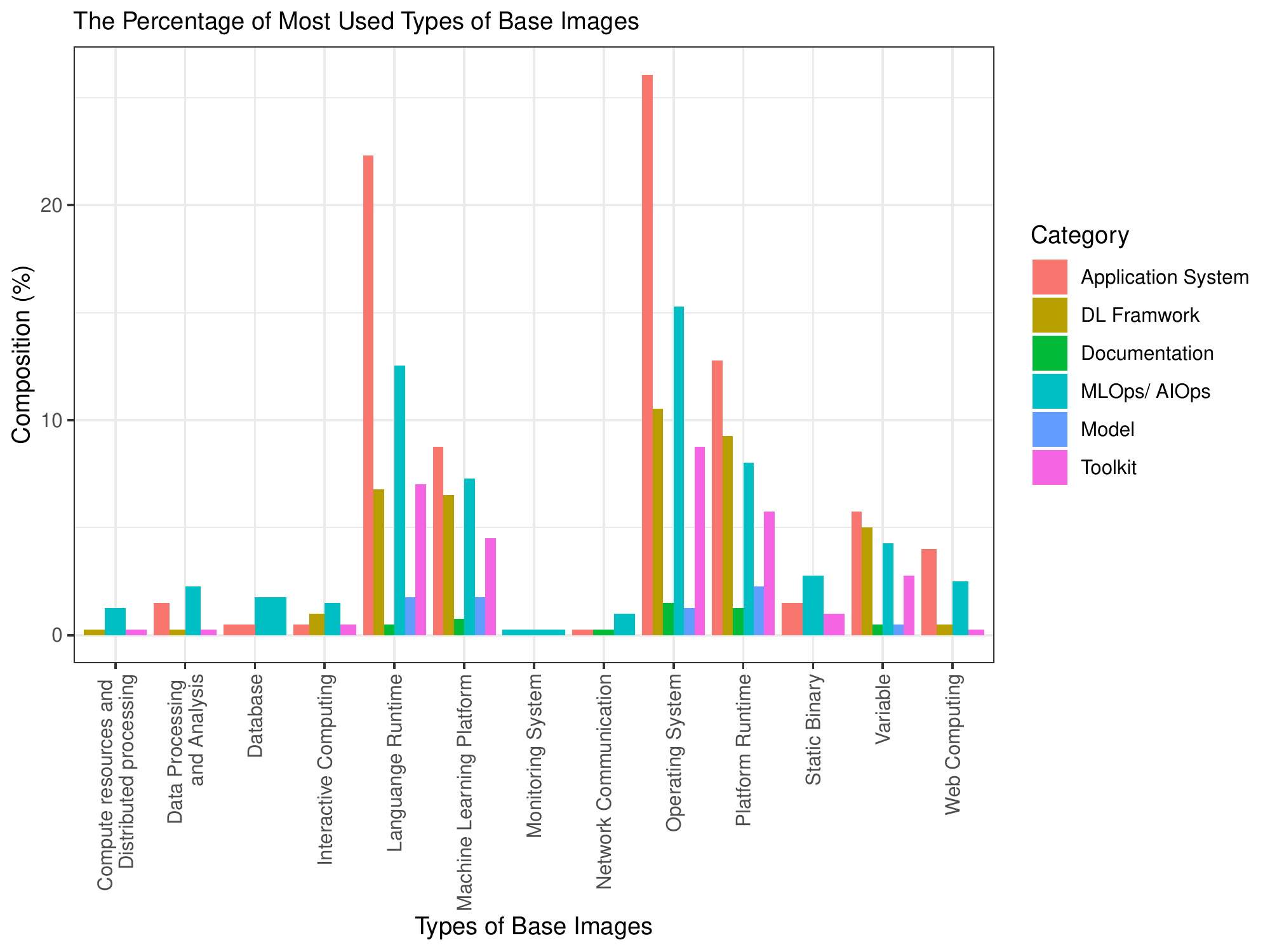}
         \caption{The percentage distribution of the category 
         of the commonly used Base Images.}
         \label{fig:Base_image_category}
     \end{subfigure}
        \caption{The percentage composition of the most used Base Docker image types for Deploying ML-based  software projects and the Base Image types.}
        \Description{The composition of the most used Base images and the image types.}
        \label{fig:BaseImages}
        \vspace{-15pt}
\end{figure}

\nd Figure~\ref{fig:Base_image} present the composition of the most used base images extracted from the Dockerfiles of the selected ML software projects. The sign `$\{\}$' indicate the placeholder of the organization name providing the  image (i.e., a non-official images - community-based image) while the sole name identifier e.g., ubuntu, node, alphine indicate the official images. As shown in Figure~\ref{fig:Base_image}, overall the official base image ubuntu 
is the most commonly used image followed by cuda and python. As their names suggest, ubuntu image hold the ubuntu operating system, while cuda 
is a  toolkit to develop, optimize and deploy GPU-accelerated applications. Other images such as tensorflow, alpine, pytorch are also among the most commonly used base images. 

\nd To further summarize the use of base images, we manually classified them into 13 different types. Figure~\ref{fig:Base_image_category} reports the composition of the base images types. The base images of type `Operating system (OS)' are the images that contain a blank operating system, without further software being installed. Type `Language runtime' images contain a runtime needed to execute applications written in a specific programming language. The `Platform Runtime' runs at the Platform Runtime layer on top of OS and Language runtime, allowing the software to use certain computing platform types (e.g., Intel CPU, GPU) of an organization's infrastructure for high availability, performance, security, and reliability. The rest of the categories are the modified versions of base image (also called application type) usually from the official image to suite the specific need. For instance, the `Machine learning platforms' are specific application type base images bundled with different tools, including DL frameworks, AutoML, or online platforms (e.g., Azure ML, AWS Sagmaker) to streamline the development, training, deployment, or managing of machine learning workflow. Similarly, the `Database' base images contains database such as mongo, postgre.  Specifications, a label `Variable' is used for the base images that are specified using placeholders for parameters (e.g.,  $\{namespace\}/\{prefix\}:\{tag\_id\}$) to be filled out dynamically during the running or building of the image. 

\begin{center}
\fbox{\parbox{0.45\textwidth}{
The most commonly used base images for building the Docker images for ML-based software projects are related to Operating systems, Platform Runtime, Language Runtime, and DL frameworks. This implies that ML engineers use Docker images mostly to help with the platform portability 
such as transferring the software across the operating systems, runtime such as CPU, GPU usage, and language constraints.
}}
\end{center}
\subsection{\textbf{RQ4: Characteristics of the Docker Images}}\label{rq4}

\begin{table*}[t]
    \centering
    \tiny
    \caption{Summary of the Docker images for deploying ML software projects. \emph{Category:} ML software projects categories, \emph{Conf. size:} The config size is the median size of the configuration as indicated in the manifest files. \emph{Arch:} the target architecture used by the Docker images. \emph{OS:} Operating system used by the Docker image. \emph{Env:} The median number of environment variables defined in the Docker images, \emph{Layers:} The median of number of main layers within the Docker images, \emph{RootFS:} The median number of Root Filesystem layers in the Docker images.
    \emph{File:} median of total number of files contained in the Docker image layers.
    \emph{Max.Level:} Median of Maximum level of sub-directory in the image layers.}
    \label{tab:image-summary}
    \resizebox{\textwidth}{!}{\begin{tabular}{m{2.2cm} | m{1.2cm} m{1.2cm} m{1.2cm} r r  r r r}\toprule
    
     \rowcolor{gray!15}
    \textbf{Category}&	\textbf{Conf. size}&	\textbf{Arch}&	\textbf{OS}&	\textbf{Env }&	\textbf{Layers}&	\textbf{RootFS}&\textbf{Files}&\textbf{Max.Level}\\\midrule
    Model&	13.74&	amd64&	linux&	9&	19.5&	22&	51,943&	17\\
     
 \rowcolor{gray!10}
Application System	&11.36&	arm, arm64, amd64&	linux	&	8&	17&	19&	62,660&	16\\

 MLOps/ AIOps&	11.7&	amd64&	linux&	8&	13&	17&	49,768&	14\\

 \rowcolor{gray!10}
 Toolkit&	11.38&	s390x, amd64&	windows, linux&	8&	17&	19.5&	61,828&	17\\

 DL Framwork&	12.45&	amd64&	windows, linux&	10&	16&	19.5&	62,967&	17\\
 
  \rowcolor{gray!10}
 Documentation&	8.56&	amd64&	linux&	5.5	&12&	13&	89,581&	18\\
\bottomrule
    
    \end{tabular}}

\end{table*}

Table~\ref{tab:image-summary} reports the general summary of the Docker images (such as the configuration size of the images, the median number of layers, the target hardware platform architecture and operating systems, and the median number of environment variables) extracted from the image manifest and configuration files of the selected ML software projects. The results in Table~\ref{tab:image-summary} indicate that all the images target the Linux operating system and amd64 Linux kernel architecture. The Docker images for ToolKit and Application System categories target multiple architecture (multi-arch) platforms (i.e., arm, amd64, and s390x). The s390x hardware architecture employs a channel I/O subsystem in the System/360 tradition. The images built on arm-based architecture target the portable devices running Edge and IoT hardware such as ML-based software projects running on Raspberry Pi (e.g., an autonomous racing robot Application System:~\texttt{`sergionr2/RacingRobot'}~\footnote{\url{https://github.com/sergionr2/RacingRobot}} extracted from the list of studied ML software projects). 

\nd Table~\ref{tab:image-summary} shows that most of the ML-based software projects have the median size of the image configuration files of about 11MB, and the number of files contained in the image layers are more than 50,000 in 90\% of the studied ML software projects. We also observed from Table~\ref{tab:image-summary} that most of the layers have more than 16 maximum sub-directories. These results are an indication that the majority of the images for deploying ML software have large number of files containing deeply nested directories. Contrary to traditional software projects, Zhao, Nannan, et al.~\cite{zhao2019large} observed that majority of the Docker images layers consist of small number of files and the directory hierarchy are not deeply nested. Consequently, our results implies that more resources are required to run the Docker images for building ML based software projects.

\begin{center}
\fbox{\parbox{0.45\textwidth}{
Due to the numerous desirable features of Docker technology, likewise the images for deploying ML software projects have numerous desirable characteristics such as targeting different operating systems and hardware architecture platforms. Moreover, ML engineers tend to use multiple different configuration settings on the images, such as environment variables more than five dynamic-named values that are queried during the running processes of the ML software projects. Subsequently, the images for deploying ML software projects tend to have larger files with deeply nested directories that are likely to introduce the computation overheat. 
}}
\end{center}

\nd According to Table~\ref{tab:image-summary} it's yet surprising to see that the Documentation category with the relatively smaller size of configuration, fewer environment or layers instead contains 37,638 more files than the Model category with a larger configuration size. This motivate us to investigate the different types of files contained in the layers and their respective capacity (size). 
\subsubsection{\textbf{Composition of the File in layers and their size}}

\nd Here we are interested in characterizing the files contained in the image layers in terms of their size and types. As the initial step to understanding the contained files in the layers, we observed that only a small percentage of file types are the most used files within the image layers by analyzing the whole dataset. For example, in the MLOps/ AIOps category, only 3\% (135) different file types (e.g., files related to script/ source code, Executable, Object code, Libraries including the ELF/COFF, or files associated with Python local packages storage (site-packages directory)) take up 98\% as the most occurring types of files. 

\begin{figure}
     \centering
     \begin{subfigure}[b]{\linewidth}
         \centering
         \includegraphics[width=\linewidth]{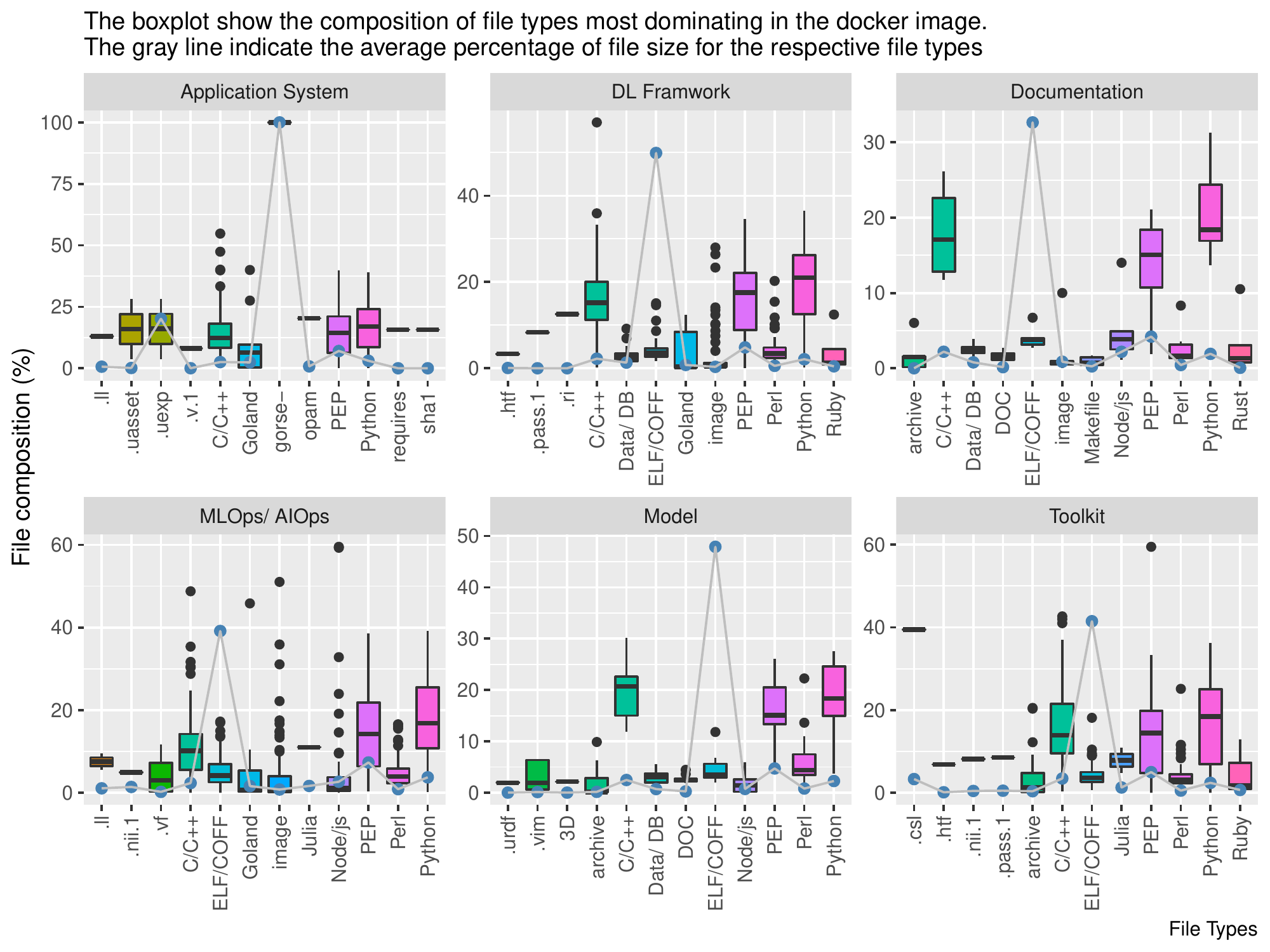}
         \caption{The mean percentage composition of the commonly used file types in the image layers and their corresponding percentage memory usage.}
         \label{fig:file-composition}
     \end{subfigure}
     \hfill
     \begin{subfigure}[b]{\linewidth}
         \centering
         \includegraphics[width=\linewidth]{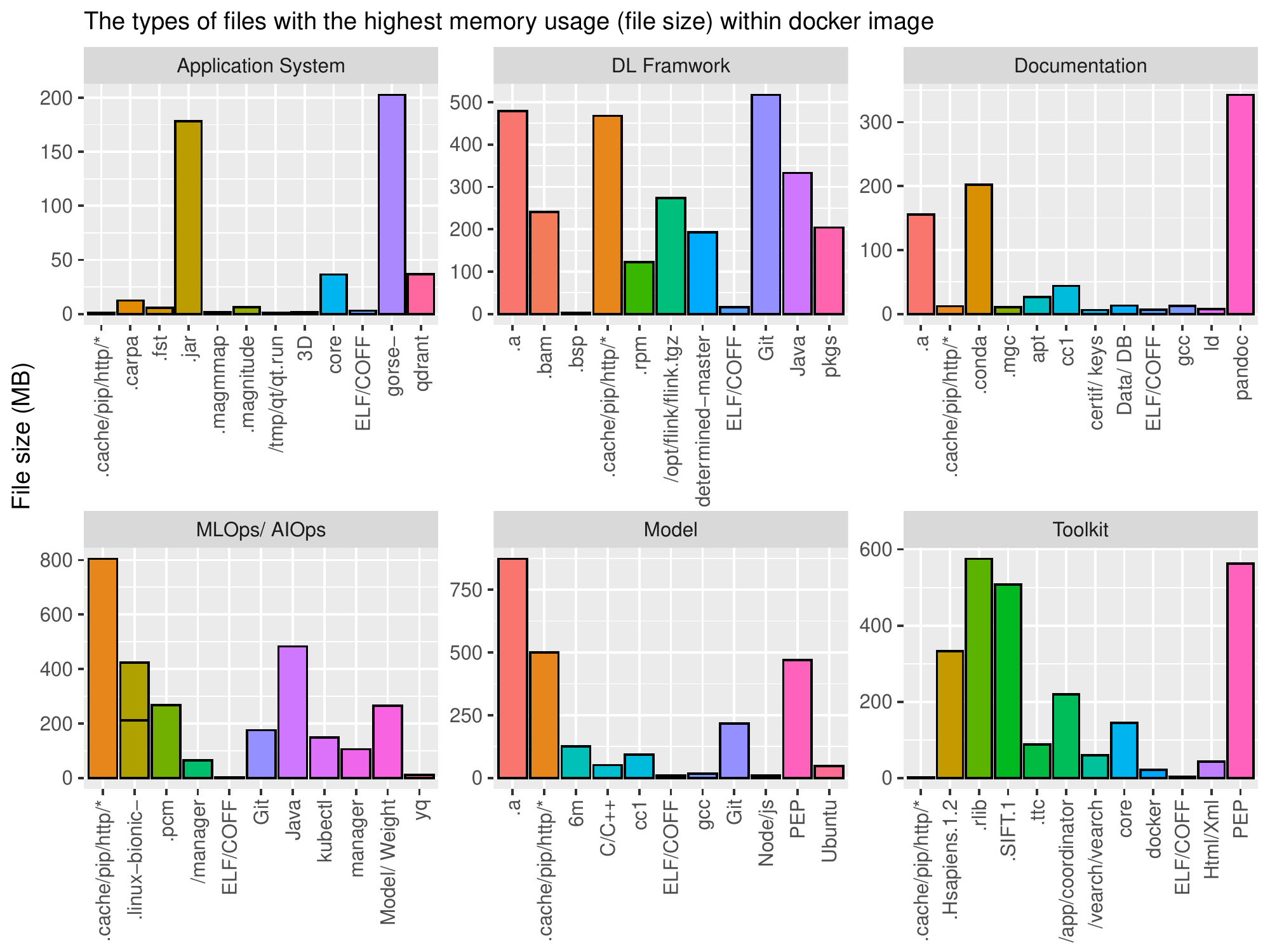}
         \caption{The types of files that take the highest memory (file size) within the Docker images.}
         \label{fig:File-size-type}
     \end{subfigure}
        \caption{The summary statistics of the commonly used file types in Figure \ref{fig:file-composition} and files types with highest memory usage is shown in \ref{fig:File-size-type}}
        \Description{The summary statistics of the commonly used file types (i.e., file types composition) in Figure \ref{fig:file-composition} and files types with highest memory usage is shown in \ref{fig:File-size-type}}
        \label{fig:File-composition-and-size}
        \vspace{-15pt}
\end{figure}

\nd In Figure~\ref{fig:File-composition-and-size} we report the $12$ most commonly used file types (in Figure~\ref{fig:file-composition}) and the files with the largest size (~\ref{fig:File-size-type}). We categorised the files basing on the file extension and their purpose following the idea of the related works~\cite{zhao2019large}. Notably, in Figure~\ref{fig:File-size-type} we show the composition of the commonly used file types as a Boxplot and their corresponding percentage memory usage (median values) as the line plots.

\nd To highlight the file types, the `.a' files represent the static libraries used to perform linking during the creation of an executable or another object file. `Data/ DB' are the data files (e.g., csv, json, `.dat', `.npy', `.log') and Database related storage (e.g., `sqlite', `.db', `.dbf', `.mdb', `.pdb') usually used as input or output. `archive' are the archived data, e.g., `.zip', `.gzip', `.tar', `.xz', `.bzip', `.bz2'. `Image' label are images e.g., `.png', `.jpeg', `.svg', `.fits', `.tiff', `.eps', `.gif', `.pic', `.ico', `.jpg', `.xpm', `.pgm'.  `Model/ Weights' include models or model weights related files saved using the popular format e.g., `.pb', `.hdf5', `.pkl', `.mlmodel', `.ckpt', `.onnx'. `PEP' in Figure~\ref{fig:File-composition-and-size} indicate the Python packages stored after a manual build and installation of the Python source code (contained in site-package or dist-packages directories) that do not fall under the rest of the categories. Similarly, for the files that we could not directly identify the types (without extension), we represented them using the path information, e.g., `.cache/**';- representing the cached data stored to speed up the performance

From Figure~\ref{fig:File-size-type} we can clearly observe that the files related to Python script and C/C++ source code are the most dominating file types contained in the image layers. The adoption of Python and C/C++ programming languages in ML software projects is not surprising due to their general-purpose nature, making them used in research, development, and production at small and large scales (e.g., Google and YouTube). For instance, Python provides unique features that contain an extensive and comprehensive library for scientific computation and data analysis and offers dynamic systems and automatic memory management, with extensive libraries for scientific computation and data analysis. However, the high proportion of Python related files subsequently leads multiple Python packages across layers shown by the high proportion of Python packages indicate by `PEP' files. 
These files (Python, C/C++) however do not take as much memory space as some less commonly used types of files, such as ELF/ COFF or gorse~\cite{Gorse:documentation} files.

In Figure~\ref{fig:File-size-type} we report the top $12$ types of files with the largest size (computed by taking the median of the file size of each file type in a category) in the image layers of the studied ML software project categories. We can see that different file types contribute to the larger image size across the categories of the ML software projects. For example, `.a' files have the largest size in the Model category, while .jar and gorse- files indicate the largest median file size in the Application System category. 


\begin{center}
\fbox{\parbox{0.45\textwidth}{
On the one hand, the image layers are composed of script and source code related files mostly written in Python and C/C++. On the other hand, these files do not occupy as much memory space as the files related to Executable, Object code, Libraries (e.g., ELF/COFF, debian libraries). 
Future works may propose an efficient way to store such files (e.g., using efficient compression) 
to avoid computation overhead.  
}}
\end{center}


\section{Related Works}\label{sec:related-works}

In this section, we discuss prior works related to our study.  
The study related to Docker and its metadata has recently gained much attention from the research community. Zhao et al.~\cite{zhao2019large} carried out a large-scale analysis of container images stored in Docker Hub. They presented an approach to exhaustively crawl and efficiently download Docker images from the Docker Hub. They analyzed  1,792,609 layers and 5,278,465,130 files accounting to 47 TB Docker dataset. Specifically, their analysis focused on various metrics related to layers, images, and files. For example, they studied image and layer sizes, deduplication ratio, Docker image compression, and image popularity metrics. Moreover, they revealed that layers are rarely shared between Docker images which increases storage utilization. Also, they indicate that file-level deduplication can eliminate 96.8\% of the files. Skourtis et al.~\cite{skourtis2019carving} studied the deduplication ratio of 10,000 most famous Docker images in Docker Hub to motivate the future approach 
to more efficient organization of Docker images. Shu et al.~\cite{shu2017study} studied 356,218 Docker images to understand the security vulnerabilities in Docker Hub. They reported a strong need for more automated and systematic methods of applying security updates to Docker images. While the number of images is similar to the studied images by Zhao et al.~\cite{zhao2019large}, Shu et al. specifically focused on a subset of 100,000 repositories and different image tags in these repositories. Cito et al.~\cite{cito2017empirical} conducted an empirical study for characterizing the Docker ecosystem, focusing on prevalent quality issues and the evolution of Dockerfiles based on a dataset of 70,000 Dockerfiles.  They reported that most quality issues (28.6\%) arise from missing version pinning (i.e., specifying a concrete version for dependencies). Also, they indicated that 34\% of Dockerfiles could not build from a representative sample of 560 projects. They believe that integrating quality checks could result in more reproducible builds. For example, quality checks to issue version pinning warnings into the container build process. Moreover, they reported that most popular projects change more often than the rest of the Docker population, with 5.81 revisions per year and 5 lines of code changed on average. Most changes deal with dependencies that are currently stored in a relatively unstructured manner.  However, in their study, they did not focus on actual image data. Slacker~\cite{harter2016slacker} investigated 57 images from Docker Hub for various metrics and used the results to derive a benchmark from evaluating the pull, push, and run performance of Docker graph drivers based on the studied images. Brogi et al.~\cite{brogi2017Dockerfinder} proposed Dockerfinder, a microservice-based prototype used to search for images based on multiple attributes, e.g., image name, image size, or supported software distributions. It also crawls images from a remote Docker registry, but the authors do not describe their crawling mechanism. 

\nd Different from these works that study Docker practices in general software projects, our work studies the practices of using Docker in the deployment of ML projects which is different from the deployment of other traditional software (e.g., considering the deployment of a models).

\section{Threats to Validity}\label{sec:threats}

\textit{Internal Validity}: This study includes a significant amount of manual labeling which may lead to subjective results. To mitigate this threat, two authors of this paper did multi-round labeling in parallel followed by discussions to improve the consensus of the results. 
For example, we did 50\% of the labeling in parallel, then reviewed and discussed the labelling results. Once we agreed upon the results, we continued up to 75\%, followed by discussions and reviews, until finally we reached 100\% of the labelling. 

\textit{External Validity}: 
To study the deployment practice of ML-based  software projects on Docker, we analyzed 406 
open-source ML-based  software projects from GitHub. Our selected projects are related to different domains of machine learning. Our results can be considered as a reference for ML engineers and researchers to improve their understanding of Docker deployment in the context of ML-based software projects. However, our results may not generalize to all ML-based software projects. We shared our dataset online \cite{moses_openja_2022_6461319}. Future studies are welcome to replicate and validate our work in other types of ML-based software projects.


\section{Discussion and Conclusion}
\label{sec:conclusion}

As our study's initial step, we categorised ML-based projects using Docker for deployment into six (6) categories, including ML Applications, MLOps/ AIOps, Toolkits, DL Frameworks, Models, and Documentation. Indeed, the adoption of Docker  
in these variety of ML software projects is not surprising given the numerous advantages of using Docker, such as its lightweight and fast start time compared to other related virtualization technology. Next, we derived a taxonomy of 21 major categories representing the purposes of using Docker. 
We believe that ML engineers can learn about using Docker 
for the similar purposes in their future deployment of ML-based software projects. Moreover, we showed that ML engineers use Docker images mostly to help with the platform portability, such as transferring the software across the operating systems, runtimes such as CPU, GPU usage, and language constraints. However, we also found that more resources may be required to run the Docker images for building ML-based software projects due to the large number of files contained in the image layers with deeply nested directories. 

As the first study on the use of Docker for the deployment ML-based software projects, our work provide insights for ML practitioners in their future deployment of their projects (e.g., learning from the deployment purposes of other ML-based projects) and for containerization providers (e.g., Docker) to improve their services for ML-based projects (e.g., improve the storage efficiency for specific types of containerized files).

Similarly, we recommend that future researchers investigate each of the $21$ topics in detail and identify the challenges of using docker in deploying ML-based projects effectively for such purposes. Also, due to the high percentage of Filesystem and dependencies related commands, we encourage the researchers to investigate and propose an efficient procedure for defining the dependencies and Filesystems to help minimize the complexity of Dockerfile and improve their maintainability. Finally, we observe that a small set of the contained files occupy relatively large image space, which calls for future works to propose more efficient approaches (e.g., using efficient compression) to store these large files related to Executable, Object code, Libraries (e.g., ELF/COFF, debian libraries).

\begin{acks}
This work is funded by the Fonds de Recherche du Québec (FRQ), Natural Sciences and Engineering Research Council of Canada (NSERC), and Canadian Institute for Advanced Research (CIFAR).
\end{acks}

\balance
\bibliographystyle{ACM-Reference-Format}
\bibliography{references}

\end{document}